\documentclass[journal]{IEEEtran}
\usepackage{graphicx}

\begin{document}
\title{Performance of Edgeless Silicon Pixel Sensors on p-type substrate for the ATLAS High-Luminosity Upgrade}

\author{Marco~Bomben, 
Alvise~Bagolini, Maurizio~Boscardin, Luciano~Bosisio, Giovanni~Calderini, Jacques~Chauveau, Audrey~Ducourthial, Gabriele~Giacomini, Giovanni~Marchiori and~Nicola~Zorzi% <-this % stops a space
\thanks{Manuscript received November 23, 2015.}% <-this % stops a space
\thanks{M.~Bomben (corresponding author) and Audrey~Ducourthial are with Laboratoire de Physique Nucleaire et de Hautes \'Energies (LPNHE) and Universit\'e Paris Diderot-Paris 7, 75252 PARIS CEDEX 05, France (e-mail: marco.bomben@cern.ch).}%
\thanks{A.~Bagolini, M.~Boscardini and N.~Zorzi are with Fondazione Bruno Kessler, Centro per i Materiali e i Microsistemi (FBK-CMM), 38123 Povo di Trento (TN),
Italy.}%
\thanks{G.~Giacomini is now with Brookhaven National Laboratory, Instrumentation Division 535B, Upton, NY - USA; was with Fondazione Bruno Kessler, Centro per i Materiali e i Microsistemi (FBK-CMM), 38123 Povo di Trento (TN), Italy} 

\thanks{L.~Bosisio is with Universit\`a di Trieste, Dipartimento di Fisica and INFN, 34127 Trieste, Italy.}%
\thanks{G.~Calderini is with Laboratoire de Physique Nucleaire et de Hautes \'Energies (LPNHE), 75252 PARIS CEDEX 05, France,  and Dipartimento di Fisica E. Fermi, Universit\`a di Pisa, and INFN Sez. di Pisa, 56127 Pisa, Italy}%
\thanks{J.~Chauveau and G.~Marchiori are with Laboratoire de Physique Nucleaire et de Hautes \'Energies (LPNHE), 75252 PARIS CEDEX 05, France.}%
%\thanks{A.~La Rosa is with Now at Max-Planck-Institut f\"ur Physik, F\"hringer Ring 6, D-80805 M\"nchen, Germany; was with Section de Physique (DPNC), Universit\`e de Gen\`eve, CH-1211 Gen\`eve 4, Switzerland}%
}

\maketitle
\pagestyle{empty}
\thispagestyle{empty}

\begin{abstract}
\indent In view of the LHC upgrade phases towards the High Luminosity LHC (HL-LHC), the ATLAS experiment plans to upgrade the Inner Detector with an all-silicon system. 
The n-on-p silicon technology is a promising candidate to achieve a large area instrumented with pixel sensors, since it is  radiation hard and cost effective.\\
\indent The paper reports on the performance of novel n-on-p edgeless planar pixel sensors produced by FBK-CMM, making use of the active trench for the reduction of the dead area at the periphery of the device. After discussing the sensor technology an overview of the first beam test results will be given.
\end{abstract}


\section{Introduction}
\IEEEPARstart{T}{he} 
 ATLAS collaboration will upgrade the current Pixel Detector~\cite{AtlasPixels} in two phases. 
A first upgrade has already been realised during the shut-down in 2013-14, by inserting a fourth detection layer (Insertable B-Layer - IBL~\cite{IBLTDR} ) at a radius of 3.2\,cm from the beam line. 
Beyond 2023, the Phase-II luminosity upgrade for the LHC,  aims to increase the instantaneous luminosity to $5\times10^{34}$\,cm$^{-2}$\,s$^{-1}$, posing a serious challenge to 
the technology for the ATLAS tracker in the High Luminosity era (HL-LHC): the lifetime fluence for the innermost layer, including safety factors, is estimated to be on the order of
$2\times10^{16}$\,n$_{\rm eq}$/cm$^2$~\cite{HL-LHC}. Hence, in view of a possible pixel system replacement after 2023, new pixel sensors are under study.   
Within the Planar Pixel Sensor collaboration (PPS)~\cite{bib:PPS} several optimizations of the well-known silicon planar technology are under investigation.

The new pixel sensors will not only have to sustain the harsher environment, but also have to show high geometrical acceptance without overlapping
 adjacent modules. Hence the inactive area has to be reduced significantly.  
One way to reduce or even eliminate the insensitive region along the device periphery is offered by
 the ``active edge'' technique~\cite{3dKenney}, in which a deep vertical trench is etched along the device periphery throughout the entire wafer thickness, 
thus performing a damage free cut (this requires using a support wafer, to prevent the individual chips from getting loose). 
The trench is  then heavily doped, extending the ohmic back-contact to the lateral sides of the device: the depletion region can then extend to the edge without causing 
a large current increase.
 
The active edge technology  has been chosen for a first production of  n-on-p planar sensors at FBK. 
Studies performed with TCAD simulation tools helped in defining the layout  and making a first estimation of the charge collection efficiency 
expected after irradiation~\cite{bib:nim2012}.

\begin{figure}[h!]
\centering
\includegraphics[width=3.5in]{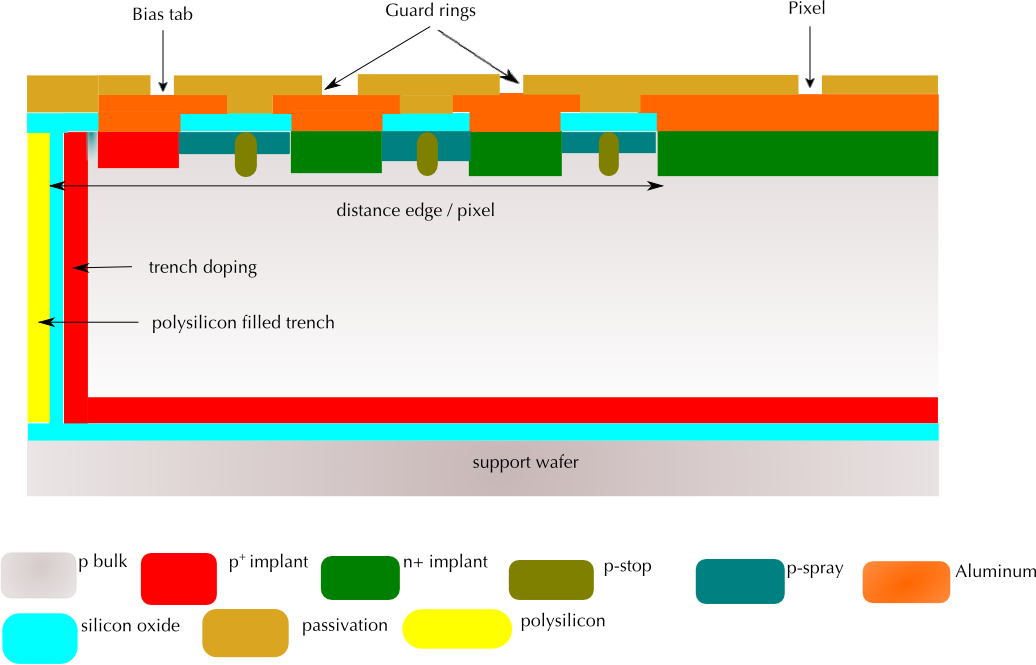}
\caption{Sketch of the pixel sensor (edge region).}
\label{fig:sensor}
\end{figure}

\indent In this paper  an overview of the performance of novel edgeless silicon n-on-p planar pixel sensors fabricated at FBK-CMM \cite{bib:nim2012} is presented. 
The sensors have been produced on 4" wafer of high resistivity Float zone (Fz) material, with a thickness of $200\,\mu m$, by making use of the active trench concept \cite{3dKenney}. This technology requires a single-sided process, featuring a doped trench, extending all the way through the wafer, and completely surrounding the sensor (see Fig.\ref{fig:sensor}) The presence of a support wafer was then required.
\\The wafer layout contains nine ATLAS FE-I4 \cite{FEI4} compatible pixel sensors with different number of guard-rings - GRs - (0, 1, 2, 3, 5 and 10) and different pixel-to-trench distances (100, 200, 300 and 400\,$\mu$m) in order to evaluate the best sensor configuration (see Table\,\ref{tab:layout_split}). In addition, several test-structures have been also implemented to study the electrical behaviour of an even larger number of GRs and pixel to trench distances combination.
\begin{table}[!ht]
\begin{center}
\begin{tabular}{cc}

\# of GRs & pixel-to-trench distance (${\rm \mu m}$) \\
\hline
0 & 100 \\
1 & 100 \\ 
2 &100\\
3 & 200 \\ 
5 & 300 \\ 
10 & 400 
\end{tabular}
\end{center}
\caption{\label{tab:layout_split}List of FE-I4 sensors layout. Two different designs have been included for the sensor with 3 GRs and 200~$\mu$m pixel-to-trench distance.}
\end{table}  

\begin{figure}[!t]
\centering
\includegraphics[width=3.5in]{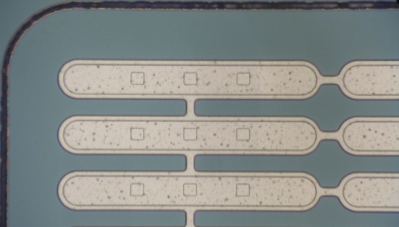}
\caption{Edge area of a test structure featuring 100~$\mu$m pixel-to-trench 
distance and no GRs; the black line is the doped trench.}
\label{fig:100um_0GRs}
\end{figure}

In Figure~\ref{fig:100um_0GRs} the edge area of a test structure 
is reported; this test structure featured FE-I4-like pixels. The black line 
sourrinding the pixels' electrodes is the doped trench; the distance between 
the trench and the pixels' electrodes is 100~$\mu$m, and, as it can be seen,
 there are no guard-rings.

Since some sensors were to be bump-bonded to  FE-I4 read-out chips, 
it was necessary to select good sensors at the wafer level, by measuring their I-V characteristics.
 For this purpose, an additional layer of metal was deposited over the passivation and patterned into stripes, each of them shorting together a row of pixels, contacted through
 the small passivation openings foreseen for the bump bonding.
This solution has already been adopted for the selection of good 3D FE-I4 sensors for the ATLAS IBL~\cite{6522814}.
After the automatic current-voltage  measurement
 on each FE-I4 sensor, the metal was removed by  wet etching, which does not affect the electrical characteristics  of the devices.

\begin{figure}[!t]
\centering
\includegraphics[width=2.75in]{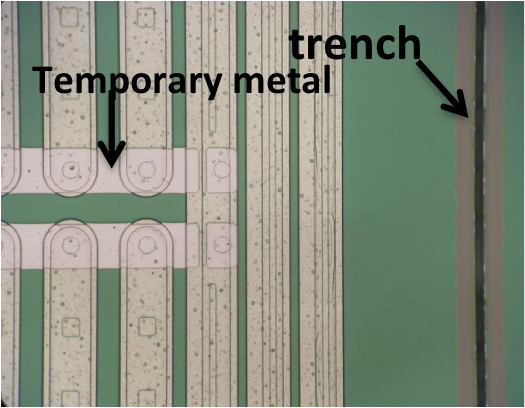}
\includegraphics[width=3.5in]{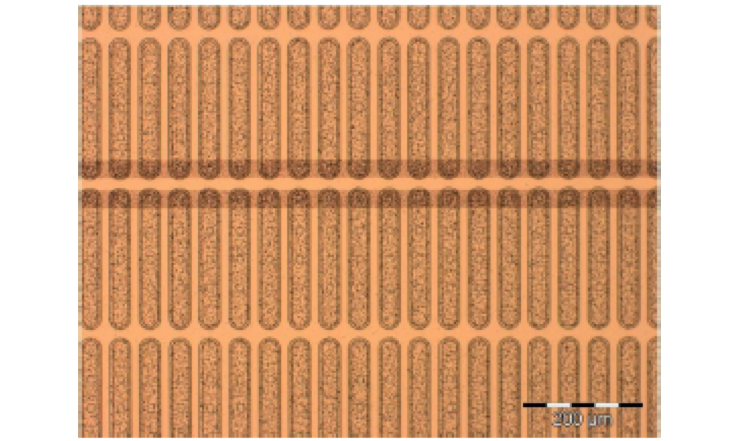}
\caption{Top) layer of temporary metal, shorting columns of pixels in a FE-I4
 compatible sensor. Bottom) the same sensor after the temporary metal removal.}
\label{fig:temp_metal}
\end{figure}

In Figure~\ref{fig:temp_metal} a sensor before and after the temporary 
metal removal is reported.

\section{2015 DESY testbeam results}
In march 2015 one FE-I4 pixel module was tested on beam at DESY~\cite{desytb}.
The module has been realized
 by bump-bonding one FE-I4B~\cite{FEI4} readout chip and one active edge pixel sensor from
 our n-on-p production; the sensor had 100~$\mu$m pixel-to-trench distance and 
no GRs; it can be said that for what concerns the edge region this 
 sensors looked as the one in Figure~\ref{fig:100um_0GRs}. 
In what follows this pixel module under test will be reffered to as LPNHE5.

In DESY the measurements were performed on beamline 21~\cite{desytb}. Electrons 
of 4~GeV/c momentum where impinging normally to our LPNHE5 module surface. 
The LPNHE5 device under test (DUT) was placed between the two arms of the DATURA 
beam telescope~\cite{ref:eudetreport200902}.

Several configurations were tested, including different bias voltage values 
for the DUT and different threshold values for the FEI4 readout chip; 
 data were recorded too when the normal to the DUT was making an angle of 15$^{\circ}$ with 
 respect to the beam axis. All measurements were performed at room temperature.

\begin{figure}[!t]
\centering
\includegraphics[width=3.5in]{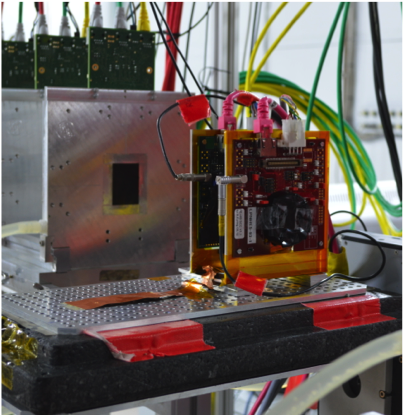}
\caption{DESY testbeam setup. Two FE-I4 pixel modules (on the right) 
were put between the two EUDET/Aida telescope arms; the first two planes 
of the downstream arm can be seen on the left. Beam was coming from the right.}
\label{fig:testbeam_setup}
\end{figure}

In Figure~\ref{fig:testbeam_setup} a picture of the setup. 

A trigger signal was generated by the coincidence of 2 scintillators pairs, 
one upstream and one downstream of the telescope. The data acquisition rate was limited to $\sim250$~Hz 
 by some data acquisition problems. 

Data were reconstructed using the eutelescope/eudaq software~\cite{eutelescope}
 and analysed with the tbmon2 package~\cite{tbmon2}.
 Tracks were built  from clusters of hits in the telescope planes. 
An iterative alignment fit procedure, followed by a tracks fit step 
 allowed to obtain a pointing resolution of about 3~$\mu$m, thanks also to the telescope planes pixel sensor small pitch
 ($\sim 18 \mu$m)~\cite{HuGuo2010480}.

Once a set of good tracks was defined, the clusters properties, the hit-efficiency and the spatial 
 resoultion of the LPNHE5 DUT were studied.

\subsection{Cluster properties}

\begin{figure}[!t]
\centering
\includegraphics[width=3.5in]{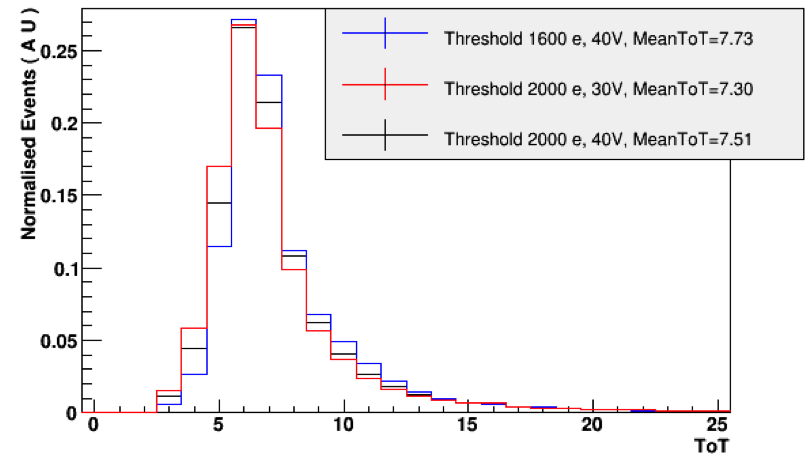}
\caption{Charge in clusters as a function of different threshold and bias voltage.}
\label{fig:cluster_charge}
\end{figure}

Clusters were formed by grouping neighboring pixels
 that fired in time with tracks registred by telescope. For tracks at normal 
 incidence about 20\% of the clusters were composed by more than one pixel; 
 for tracks at  15$^{\circ}$ more than 60\% of the clusters were composed by 
 two or more pixels.
In Figure~\ref{fig:cluster_charge} the cluster charge, measured as
 Time-over-Threshold (ToT), is shown. As expected, lower threshold and higher 
 bias voltage~\footnote{The depletion voltage was 20~V.} gives the cluster more charge.

\begin{figure}[!t]
\centering
\includegraphics[width=3.5in]{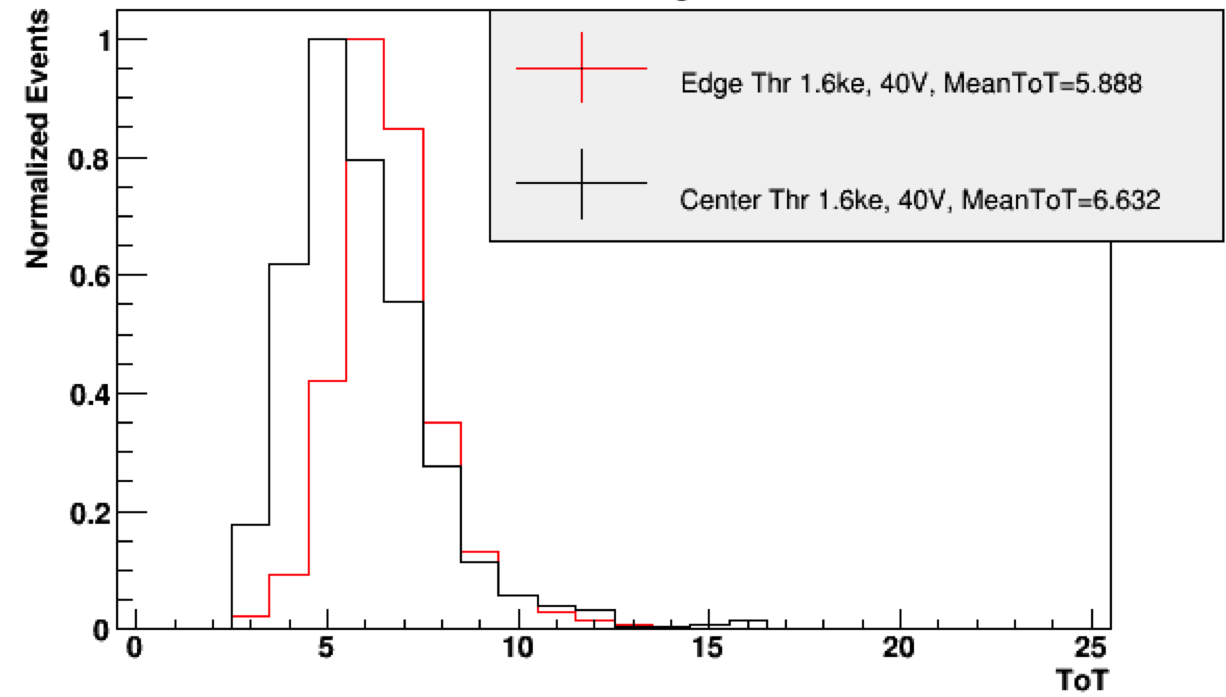}
\caption{Charge in isolated single pixels:
 a comparison between edge (red) and central (black) pixels.}
\label{fig:edge_charge_cut}
\end{figure}

Edge pixels properties were studied, first of all to check if they show the same performance as the ``central'' ones.
 In Figure~\ref{fig:edge_charge_cut} the charge measured by isolated 
single pixels is presented, 
 separately for the edge and the center zone. It can be seen that edge pixels 
 are at least as effective as the central ones in collecting charge; they are 
 actually collecting more charge on average and this can be explained by 
 charge collection from the un-instrumented area which hence is active. 
 This is a first strong indication that the active edge technology works.

\subsection{Hit-efficiency}
 The LPNHE5 hit-efficiency performance has been studied by looking at hits 
 on the DUT close to the extrapolated track impact position.
 Table~\ref{tab:eff} reports the hit
 efficiency~\footnote{hit efficiency errors are negligible} for the different 
tested  configuration.

\begin{table}[!ht]
\begin{center}
\begin{tabular}{ccc}

bias (V) & threshold (ke) & efficiency (\%) \\
\hline
30 & 2.0 & 98.4 \\
40 & 2.0 & 98.7 \\ 
40 & 1.6 & 99.1
\end{tabular}
\end{center}
\caption{\label{tab:eff}Hit-efficiency for LPNHE5 as a function
 of bias voltage and threshold.}
\end{table}  

The hit efficiency is always above~98\%; it is better than~99\% for 
the best configuration (40~V bias voltage and 1.6~ke threshold ). This is a remarkable result: even 
at moderate bias voltage the detector is full efficient.

\begin{figure}[!t]
\centering
\includegraphics[width=3.5in]{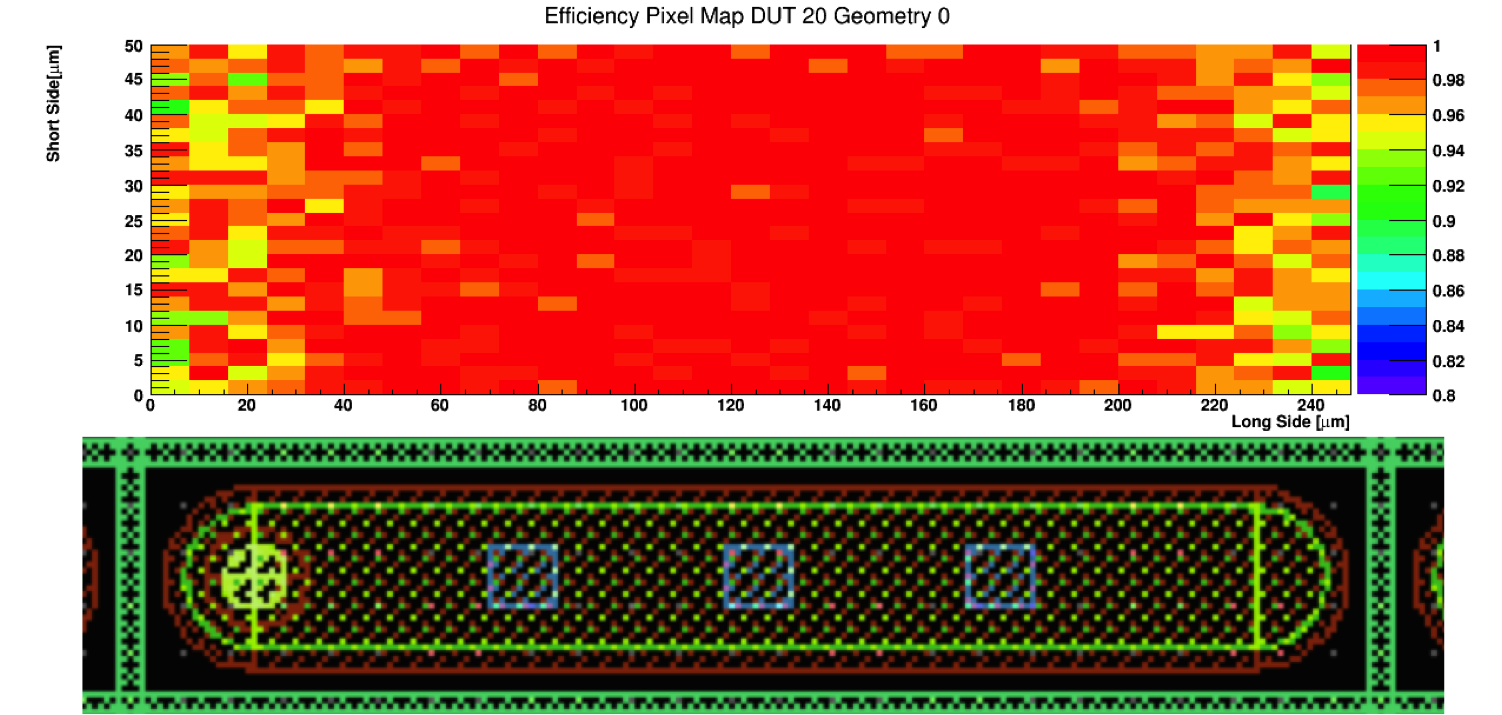}
\caption{Top) in-pixel hit efficiency. Bottom) schematic of the pixel cell.}
\label{fig:inpixel_efficiency}
\end{figure}

In Figure~\ref{fig:inpixel_efficiency} the so-called ``in-pixel'' hit
 efficiency 
is shown for 40~V bias voltage and 1.6~ke threshold. All the pixels have 
 been superimposed in order to study the hit-efficiency as a function 
 of the track impact point. 

As it can be seen the efficiency is very uniform across the pixel cell, 
above ~95\% everywhere. The fact that after the temporary metal removal 
there were no biasing structures left made the cell very effective 
 everywhere. A small less efficient region is visible
 at the corners of the cell; this is due to charge sharing with the nieghbouring cells.

\begin{figure}[!t]
\centering
\includegraphics[width=3.5in]{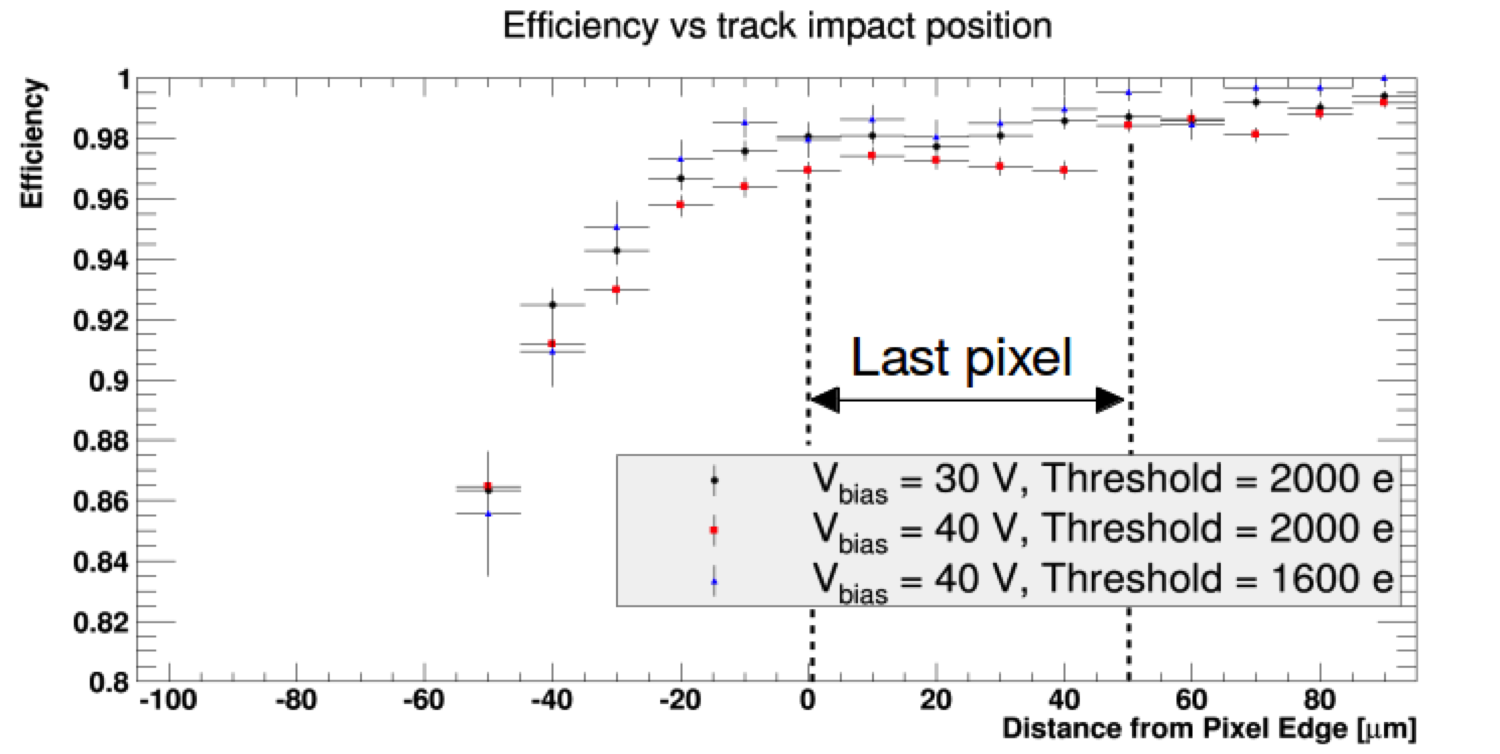}
\caption{Hit-efficiency at the sensor edge as a function of the 
 track impact position.}
\label{fig:edge_efficiency}
\end{figure}

To completely validate the active edge approach the hit-efficiency at the 
sensor edge had to be measured. The results are shown in
 Figure~\ref{fig:edge_efficiency}. The efficiency is measured as a function 
 of the track impact point for the row of pixels closest to the trench. 
Points with abscissa lower than 0 correspond to tracks passing in the 
 volume within the pixels and the trench. It can be seen that the 
 hit-efficiency is still above 90\% for all configurations even 40~$\mu$m 
 away from the last pixel implant~\footnote{there are no usable tracks 
 beyond -50~$\mu$m}. This result represents the validation of the 
 active edge approach: the un-instrumented volume is active and highly 
 efficient.  
 
\subsection{Spatial resolution}

\begin{figure}[!t]
\centering
\includegraphics[width=1.7in]{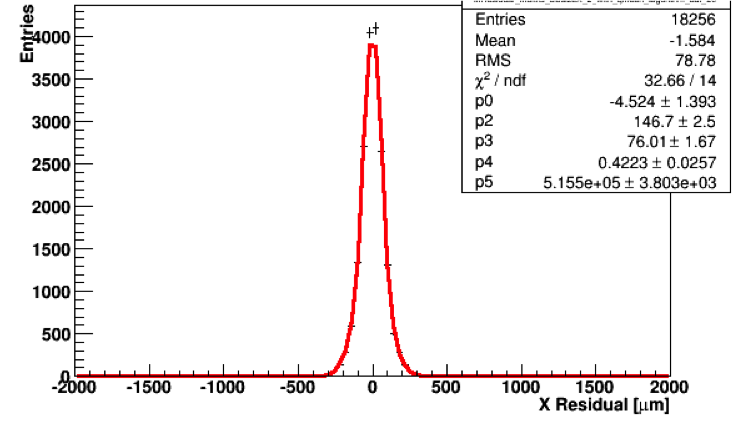}
\includegraphics[width=1.7in]{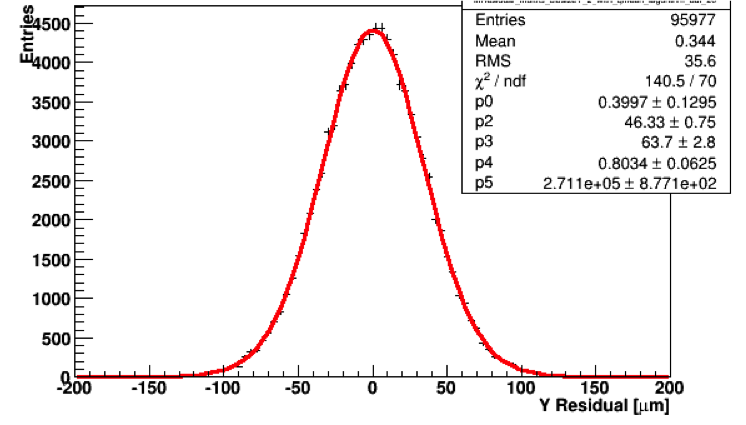}
\caption{Hit-residuals. Left) long pixel direction; right) short pixel direction.}
\label{fig:residuals}
\end{figure}

To evaluate the spatial resolution of the LPNHE5 module the hit-residuals, 
defined as the reconstructed hit position minus the track impact 
position, were studied. The hit-residual distributions for the long 
(250~$\mu$m) and short (50~$\mu$m) pixel directions are presented in 
 Figure~\ref{fig:residuals}. 
The analysis is strongly limited by the multiple scattering effect, 
estimated to be of the order of 30~$\mu$m. 
 The RMS values of the hit-residual distributions for the long and short pixel directions are $\sim 80\mu$m 
 and $\sim 34\mu$m, respectively; the results are consistent with the expectations.
 
\section{Conclusions}

In this paper we reported the performance of thin n-on-p pixels 
 aimed at the HL-LHC phase of ATLAS. These pixels are characterized 
 by their reduced un-instrumented area at the detector periphery. 
 This was possible thanks to the active edge technology. 

The testbeam results showed that our detector has a very high
 hit efficiency, very uniform across the detector.
 It was also shown that the pixels at the detector edge 
 collect charge from the un-instrumented volume as well. 
 More important, the detector is very effective also well beyond the pixels 
 area.

\section*{Acknowledgment}
We thank the ITk pixel testbeam members that contributed to make these measurements possible

Thanks to A. Macchiolo, N. Savic and S. Terzo from MPI for useful discussion.
We thank CERN for letting us use their laboratory.

 This work was supported in part by the Autonomous Province 
 of  Trento,  Project  MEMS2,  and  in  part  by  the  Italian  National  Institute  for
 Nuclear Physics (INFN).

The measurements leading to these results have been performed at the Test Beam Facility at DESY Hamburg (Germany), a member of the Helmholtz Association (HGF).


%\bibliographystyle{IEEEtran}
\bibliography{biblio}

\begin{thebibliography}{10}
\providecommand{\url}[1]{#1}
\csname url@samestyle\endcsname
\providecommand{\newblock}{\relax}
\providecommand{\bibinfo}[2]{#2}
\providecommand{\BIBentrySTDinterwordspacing}{\spaceskip=0pt\relax}
\providecommand{\BIBentryALTinterwordstretchfactor}{4}
\providecommand{\BIBentryALTinterwordspacing}{\spaceskip=\fontdimen2\font plus
\BIBentryALTinterwordstretchfactor\fontdimen3\font minus
  \fontdimen4\font\relax}
\providecommand{\BIBforeignlanguage}[2]{{%
\expandafter\ifx\csname l@#1\endcsname\relax
\typeout{** WARNING: IEEEtran.bst: No hyphenation pattern has been}%
\typeout{** loaded for the language `#1'. Using the pattern for}%
\typeout{** the default language instead.}%
\else
\language=\csname l@#1\endcsname
\fi
#2}}
\providecommand{\BIBdecl}{\relax}
\BIBdecl

\bibitem{AtlasPixels}
G.~Aad, M.~Ackers, F.~Alberti, M.~Aleppo, G.~Alimonti \emph{et~al.}, ``Atlas
  pixel detector electronics and sensors,'' \emph{JINST}, vol.~3, p. P07007,
  2008.

\bibitem{IBLTDR}
\BIBentryALTinterwordspacing
{ATLAS IBL Community}, ``Atlas insertable b-layer technical design report,''
  CERN, Tech. Rep., 2010. [Online]. Available:
  \url{http://cdsweb.cern.ch/record/1291633/files/ATLAS-TDR-019.pdf}
\BIBentrySTDinterwordspacing

\bibitem{HL-LHC}
\BIBentryALTinterwordspacing
S.~McMahon, P.~Allport, H.~Hayward, and B.~Di~Girolamo, ``{Initial Design
  Report of the ITk: Initial Design Report of the ITk},'' CERN, Geneva, Tech.
  Rep. ATL-COM-UPGRADE-2014-029, Oct 2014. [Online]. Available:
  \url{https://cds.cern.ch/record/1952548}
\BIBentrySTDinterwordspacing

\bibitem{bib:PPS}
\BIBentryALTinterwordspacing
C.~Nellist, ``Achievements of the atlas upgrade planar pixel sensors r\&d
  project,'' \emph{Journal of Instrumentation}, vol.~10, no.~01, p. C01027,
  2015. [Online]. Available:
  \url{http://stacks.iop.org/1748-0221/10/i=01/a=C01027}
\BIBentrySTDinterwordspacing

\bibitem{3dKenney}
{C~.J~Kenney {\it et al.}}, ``Results from 3-d silicon sensors with wall
  electrodes: Near-cell-edge sensitivity measurements as a preview of
  active-edge sensors,'' \emph{{IEEE Transactions on Nuclear Science, }},
  vol.~48, no.~6, pp. 2405--2410, 2001.

\bibitem{bib:nim2012}
{M.~Bomben {\it et al.}}, ``{Development of Edgeless n-on-p Planar Pixel
  Sensors for future ATLAS Upgrades},'' \emph{Nucl. Instr. and Meth. A}, vol.
  712, pp. 41--47, 2013.

\bibitem{FEI4}
{M.~Garcia-Sciveres {\it et al.}}, ``The fe-i4 pixel readout integrated
  circuit,'' \emph{Nucl. Instr. and Meth. A}, vol. 636, pp. S155--S159, 2011.

\bibitem{6522814}
G.~Giacomini, A.~Bagolini, M.~Boscardin, G.-F. Dalla~Betta, F.~Mattedi,
  M.~Povoli, E.~Vianello, and N.~Zorzi, ``Development of double-sided
  full-passing-column 3d sensors at fbk,'' \emph{Nuclear Science, IEEE
  Transactions on}, vol.~60, no.~3, pp. 2357--2366, 2013.

\bibitem{desytb}
\BIBentryALTinterwordspacing
\emph{testbeam.desy.de}. [Online]. Available: \url{testbeam.desy.de}
\BIBentrySTDinterwordspacing

\bibitem{ref:eudetreport200902}
A.~Bulgheroni, ``Results from the {EUDET} telescope with high resolution
  planes,'' Tech. Rep. EUDET-Report-2009-02, 2009.

\bibitem{eutelescope}
\BIBentryALTinterwordspacing
\emph{http://eutelescope.web.cern.ch/}. [Online]. Available:
  \url{http://eutelescope.web.cern.ch/}
\BIBentrySTDinterwordspacing

\bibitem{tbmon2}
\BIBentryALTinterwordspacing
\emph{https://bitbucket.org/TBmon2/tbmon2/overview}. [Online]. Available:
  \url{https://bitbucket.org/TBmon2/tbmon2/overview}
\BIBentrySTDinterwordspacing

\bibitem{HuGuo2010480}
C.~Hu-Guo \emph{et~al.}, ``First reticule size {MAPS} with digital output and
  integrated zero suppression for the {EUDET-JRA1} beam telescope,''
  \emph{Nucl. Instrum. Methods Phys. Rev. A}, vol. 623, no.~1, pp. 480 -- 482,
  2010, 1st International Conference on Technology and Instrumentation in
  Particle Physics.

\end{thebibliography}
\bibliographystyle{IEEEtran}

%\begin{thebibliography}{1}


%\bibitem{IEEEPDFRequirement401}
%IEEE Content Engineering, \emph{PDF Specification for IEEE Xplore}. Available: http://www.ieee.org/portal/cms\_docs/pubs/confstandards/pdfs/IEEE-PDF-SpecV401.pdf.

%\end{thebibliography}

\end{document}